\documentclass[11pt,tightenlines,eqsecnum,floats,aps,amsmath,amssymb,nofootinbib,prd,shownopacs,floatfix]{revtex4}
\usepackage{graphicx}
\usepackage{epstopdf}
\usepackage{latexsym}
\usepackage{amssymb}
\usepackage{amsmath}
\usepackage{color}
\usepackage{mathrsfs}
\usepackage{xparse}
\usepackage{float}
\usepackage{mathtools}
\usepackage[center]{subfigure}

\begin{document}
\renewcommand\arraystretch{2}
\newcommand{\bq}{\begin{equation}}
\newcommand{\eq}{\end{equation}}
\newcommand{\bqn}{\begin{eqnarray}}
\newcommand{\eqn}{\end{eqnarray}}
\newcommand{\nb}{\nonumber}
\newcommand{\lb}{\label}
\newcommand{\cb}{\color{blue}}
\newcommand{\cc}{\color{cyan}}
\newcommand{\cm}{\color{magenta}}
\newcommand{\rc}{\rho^{\scriptscriptstyle{\mathrm{I}}}_c}
\newcommand{\rd}{\rho^{\scriptscriptstyle{\mathrm{II}}}_c}
\NewDocumentCommand{\evalat}{sO{\big}mm}{%
  \IfBooleanTF{#1}
   {\mleft. #3 \mright|_{#4}}
   {#3#2|_{#4}}%
}

\newcommand{\PRL}{Phys. Rev. Lett.}
\newcommand{\PL}{Phys. Lett.}
\newcommand{\PR}{Phys. Rev.}
\newcommand{\CQG}{Class. Quantum Grav.}
\newcommand{\parallelsum}{\mathbin{\!/\mkern-5mu/\!}}

\title{Alternative effective mass functions in the modified Mukhanov-Sasaki equation of loop quantum cosmology}
\author{Bao-Fei Li$^{1,2}$}
\email{libaofei@zjut.edu.cn}
\author{Parampreet Singh$^2$}
\email{psingh@lsu.edu}
\affiliation{
$^{1}$ Institute for Theoretical Physics $\&$ Cosmology, Zhejiang University of Technology, Hangzhou, 310023, China\\
$^{2}$ Department of Physics and Astronomy, Louisiana State University, Baton Rouge, LA 70803, USA\\
}

\begin{abstract}

Modifications to the Mukhanov-Sasaki equation in loop quantum cosmology (LQC) have been phenomenologically explored using
polymerization of the connection and related variables in the classical expressions in order to capture the quantum gravity effects in cosmological perturbations which replace the classical big bang by a big bounce. Examples of this strategy include the dressed metric and the hybrid approaches whose inter-relationship at an effective level was demonstrated by the authors recently. In this manuscript, we propose a new family of the effective mass functions in the modified Mukhanov-Sasaki equation of LQC by investigating the polymerization of a particular form of the classical mass function in terms of variable $z_s$($=a\dot \phi/H$) which relates the Mukhanov-Sasaki variable with the comoving curvature perturbation. Using a generalized ansatz motivated by quantum gravity effects in the background dynamics we find  alternative effective mass functions which are distinct from those used in the dressed metric and the hybrid approaches with differences originating from the non-commutativity of the evaluation of the Poisson brackets and the polymerization procedures. The new effective mass functions acquire four correction terms in the effective potential whose exact forms are closely tied up with the ansatz used for polymerizing the inverse Hubble rate. In contrast to earlier works, one of these correction terms can in principle  produce sizable effects even when the bounce is kinetic dominated. Our investigation opens a new window to explore the phenomenological implications of a large family of effective mass functions in LQC which can potentially lead to significant departures from the dressed metric and the hybrid approaches in the bounce regime.

\end{abstract}

\maketitle

\section{Introduction}

Loop quantum cosmology (LQC) \cite{Ashtekar:2003hd,Ashtekar:2011ni,Agullo:2016tjh,Li:2023dwy} which is a non-perturbative quantization of cosmological spacetimes using techniques of loop quantum gravity (LQG) generically resolves the strong singularities encountered in various classical cosmological spacetimes
\cite{Bojowald:2001xe,Ashtekar:2006rx,Ashtekar:2006uz,Ashtekar:2006wn, Ashtekar:2007em,Singh:2009mz,Singh:2011gp,Singh:2014fsy,Saini:2017ggt,Saini:2017ipg} and provides a consistent framework to extend one of the most popular scenarios in classical cosmology, namely the inflationary paradigm, to the Planck regime \cite{Agullo:2012sh,Giesel:2020raf}. This extension is concerned with at least two aspects: a non-singular evolution of the background spacetime extending the past-incomplete inflationary phase beyond the big-bang singularity which in LQC is replaced by a big bounce and the evolution of the cosmological perturbations across the quantum background spacetimes in the Planck regime. In LQC literature, the physics of quantum background spacetime minimally coupled with an inflaton field with different inflationary potentials has been thoroughly investigated (see for eg. \cite{Singh:2006im,Ashtekar:2009mm,Mielczarek:2010bh,Ashtekar:2011rm,Gupt:2013swa,Linsefors:2013cd,Corichi:2013kua,Bonga:2015kaa,Ashtekar:2015dja,Zhu:2017jew,Li:2018fco,Li:2019ipm,Gordon:2020gel,Giesel:2020raf, Motaharfar:2021gwi}) while the investigation of the perturbation theory in LQC has culminated in four approaches, namely the dressed metric approach \cite{Agullo:2012sh,Agullo:2012fc,Agullo:2013ai}, the hybrid approach \cite{Fernandez-Mendez:2012poe,Fernandez-Mendez:2013jqa,Gomar:2014faa,Gomar:2015oea,Martinez:2016hmn,ElizagaNavascues:2020uyf}, the deformed algebra approach \cite{Bojowald:2008gz,Cailleteau:2012fy,Cailleteau:2011kr} and the separate universe approach \cite{Wilson-Ewing:2015sfx}. The third approach results in a divergence in the primordial scalar power spectrum at small scales \cite{Bolliet:2015raa,Li:2018vzr,DeSousa:2022rep} which can be severely constrained and the fourth approach is tailored to the long wavelength modes only, more interest has been attached to the first two approaches in the recent years to search possible quantum gravity signals in the CMB data.

The dressed metric approach is based on Langlois' work on Hamiltonian formulation of cosmological perturbations which heavily relies on using spatially flat gauge \cite{Langlois:1994ec}. On the other hand, the hybrid approach uses gauge-invariant treatment of Halliwell and Hawking \cite{Halliwell:1984eu}.
Although the dressed metric and the hybrid approaches have been developed independently by two separate groups and appear to be distinct, our recent work found that a very close relationship exists between them \cite{Li:2022evi}.
From the perspective of the construction of the perturbation theory, both approaches are based on the classical formulation of the linear perturbation theory in cosmology. Besides, the quantization of the classical linear perturbation theory in both the approaches is implemented in a {\emph{hybrid}} way. Specifically, the background dynamics is loop quantized in the improved dynamics (the $\bar \mu$ scheme) of LQC \cite{Ashtekar:2006wn} while the perturbative degrees of freedom are Fock quantized. Moreover, when it comes to the practical computations of the primordial power spectra, both approaches make use of the effective dynamics of LQC but ignore the back-reaction of the perturbations on the background spacetimes. These approaches lead to very similar modified Mukhanov-Sasaki equations in two approaches whose differences become manifest in the different forms of the effective mass functions.

As discussed extensively in our recent paper \cite{Li:2022evi}, at the level of the effective dynamics, the different effective mass functions in the dressed and the hybrid approaches originate from different sets of canonical variables used to Fock quantize the perturbations.  In particular, the dressed metric approach uses the Mukhanov-Sasaki variable $Q_{\vec{k}}$ while t he hybrid approach utilizes its rescaled version $\nu_{\vec{k}} = a Q_{\vec k}$. At the classical level, the use of the different variables results in the equivalent expressions of the classical mass function which only differ by a term proportional to the zeroth order Hamiltonian constraint that is vanishing on the classical dynamical trajectories. However, when the classical mass functions are polymerized in LQC, these two equivalent expressions lead to different effective mass functions due to the non-commutativity between the evaluation of the Poisson brackets and the polymerization procedure. As a result, although the effective mass functions in two approaches have the same classical limit in the low curvature regime which recovers the Mukhanov-Sasaki equation in classical cosmology, they actually exhibit distinct behavior in the bounce regime. In particular, the effective mass function in the hybrid approach turns out to be positive near the bounce while the one in the dressed metric approach is not \cite{ElizagaNavascues:2017avq,Iteanu:2022zha}. The different behavior of the effective mass functions in the bounce regime can in principle determine to which extent the quantum gravity effects can potentially affect the observations. This also provides a potential method to test the physical significance of different polymerization ansatz as well as the approach itself, especially when one explores non-Gaussianities.

As is well known, due to the diffeomorphism invariance of GR, the classical Mukhanov-Sasaki equation can be derived in different gauges, such as the spatially flat gauge, the longitudinal gauge etc., and given the subtleties in the canonical perturbations on the choice of gauge-fixing and reference fields  \cite{Giesel:2018opa, Giesel:2018tcw, Giesel:2020bht} one might think that the difference in effective mass functions in the dressed metric and the hybrid approach might have some root in the way gauge fixing is implemented along with polymerization procedure used for studying quantum gravity effects in cosmological perturbations. Indeed, the choice of different gauge fixing conditions can result in different equivalent forms of the mass functions in classical cosmology. In the hybrid approach, it has been shown that different gauges can lead to the same modified Mukhanov-Sasaki equation in effective LQC \cite{Fernandez-Mendez:2012poe}. Therefore, the difference between the effective mass functions in the hybrid and the dressed metric approaches does not originate from the use of different gauges. The fundamental reason for such a difference is owing to the different treatment of the effective potential and acceleration terms in the effective mass function. Unlike in the hybrid approach where both these terms are expressed in terms of phase space variables and then polymerized, in the dressed metric approach phase space variables are used only in the effective potential term whereas modified Raychaudhuri equation is used for the acceleration term.
These different ways of polymerization result in different effective mass functions. This difference essentially is owing to the non-commutativity between the evaluation of the Poisson brackets and the polymerization. Once the origin of the difference between two approaches is well understood, it is then possible to look for other alternative effective mass functions which result from the polymerization of the classical mass function.

In addition to the forms of the classical mass function used in the dressed metric and the hybrid approaches, there exists another form of the classical mass function which is obtained from the action of a spatially flat Friedmann-Lema\^itre-Robertson-Walker (FLRW) universe minimally coupled with a scalar field in the comoving gauge \cite{Baumann:2009ds}. This third form of the classical mass function can be expressed solely in terms of $z_s$($=a\dot \phi/H)$, with scale factor $a$, the homogeneous field $\phi$, and the Hubble rate $H$. This definition of $z_s$ relates the Mukhanov-Sasaki variable $\nu_{\vec{k}}$ to the comoving curvature perturbation $\mathcal R_{\vec{k}}$ via $\nu_{\vec{k}}= z_s \mathcal R_{\vec{k}}$. Therefore, it is a form of the classical mass function expressed completely without using phase space. In LQC, in the dressed metric and the hybrid approaches the ansatz used for the polymerization of $1/\pi^2_a$ and $1/\pi_a$ terms are equally important. Here $\pi_a$ denotes the momentum of the scale factor. In contrast to the polymerization of the classical mass functions in the dressed metric and the hybrid approaches, 
the polymerization based on $z_s$ variable relies solely on the ansatz for polymerizing the inverse Hubble rate which is proportional to  $1/\pi_a$ in the classical limit. Therefore, the resulting effective mass function from this third form of the classical mass function can be quite different from those already found in the dressed metric and the hybrid approaches. Our investigation confirms this observation. We find that the polymerization of this third mass function can result in alternative effective mass functions that have so far been never encountered in the existing approaches.

Let us contrast the properties of effective mass function we find in this manuscript using the variable $z_s$ with earlier approaches in LQC.
The effective mass functions in either the dressed metric and the hybrid approaches consist of linear and quadratic terms of the energy density plus an effective potential which depends on the inflationary potential and its derivatives (see Eqs. (\ref{dressed_mass})-(\ref{hybrid_mass}) in Sec. \ref{sec:eff_mass} for their exact expressions). The new effective mass function acquires four correction terms in the effective potential whose expressions solely depend on the ansatz for polymerizing the inverse Hubble rate. Besides, one of the four correction terms can also contribute dominantly in the bounce regime even when the bounce is dictated by the kinetic energy of the scalar field. Since there is an ambiguity in polymerizing the inverse Hubble rate, different ansatz can in general lead to distinctive effective mass functions. Therefore, we make use of a general ansatz parameterized by a general function $f(\rho)$ which only needs to satisfy a few general conditions. These conditions require that the  resulting effective mass functions should be finite during the entire evolution of the universe, and that the classical mass functions must be recovered in the low curvature limit. The above two conditions highly restrict the form of the function $f(\rho)$ and its asymptotic behavior. In this way, the new effective mass functions are expressed in terms of $f(\rho)$ which can be fixed in any specific polymerization ansatz. As concrete examples, we have studied in turn the consequences of the polymerization ansatz of the inverse Hubble rate effectively employed in the dressed metric approach and the hybrid approach and find that the former can lead to a well-behaved new effective mass function while the latter can only yield an effective mass function which blows up right at the bounce. However, neither of these effective mass functions result in correction terms which still contribute in the bouncing regime. Our approach allows one to explore polymerizations which allow significant contributions from the bounce regime. For this,
we find a tentative phenomenological polymerization of the inverse Hubble rate which yield an alternative effective mass function that contains correction terms which still contribute in the bouncing regime. Our investigation shows by concrete examples that even though the dressed metric and the hybrid approaches are two of the most popular perturbation theories in LQC, there still exist viable phenomenological choices which can potentially lead to distinct signatures in the bounce regime.

This manuscript is organized as follows. In Sec. \ref{sec:overview}, we briefly review the three equivalent forms of the classical mass function in the Mukhanov-Sasaki equation of linear perturbation theory in cosmology, one of which forms the starting point of dressed metric approach using $Q_{\vec k}$ variable, another for the hybrid approach using $\nu_{\vec k}$ variable and the third one for the new effective mass studied here using $z_s$ variable. Note that in spite of their different expressions, they are all actually equivalent on the classical dynamical trajectories. In Sec. \ref{sec:new_mass_function}, we analyze the polymerization of the classical mass functions. In particular, we  re-emphasize that the first two forms can give rise to the effective mass functions used in the dressed metric and the hybrid approach while the third form after polymerization yields new alternative mass function. We then  adopt a general ansatz for polymerizing the inverse Hubble rate and obtain the correction terms in the new mass function as compared with those used in the dressed metric approach. Then we specialize to three possible options to polymerize the inverse Hubble rate when using the $z_s$ variable. Here the first two ansatz are adopted respectively in the dressed metric and the hybrid approach and discuss the viability of the resulting effective mass functions. Finally, we summarize our main findings in Sec. \ref{sec:summary}.
In the following, we use the Planck units $\hbar=c=1$ and keep Newton's constant $G$ implicitly in  $\kappa$ with $\kappa=8 \pi G$.

\section{The equivalent forms of the classical mass function in the  Mukhanov-Sasaki equation}
\lb{sec:overview}

In this section, we briefly review the classical Mukhanov-Sasaki (MS) equation in the linear  perturbation theory of cosmology with an emphasis on presenting three equivalent forms of the mass function in the Mukhanov-Sasaki equation. Although these mass functions exhibit no difference on the trajectories of the classical dynamics, after polymerization they can show potential differences in the bounce regime of LQC. In the following, we present three equivalent forms of the classical mass functions that are frequently encountered in the literature.

Here we only consider the scalar sector of the classical phase space of a spatially flat FLRW universe filled with a massive scalar field which consists of four homogeneous degrees of freedom that can be parameterized by $(a,\pi_a,  \phi, \pi_\phi)$, with $a$ and $\pi_a$ standing for the scale factor and its conjugate momentum and $ \phi$, $\pi_\phi$ for the homogeneous part of the scalar field and its conjugate momentum. Following the approach used in our previous paper \cite{Li:2022evi}, assuming a non-compact $R^3$ topology of the spatial sections, we choose a fiducial cell with the volume $V_o$ to formulate the perturbation theory in the Hamiltonian formalism in which the zeroth-order homogeneous degrees of freedom satisfy the background Hamiltonian constraint \cite{Li:2022evi}
\bq
\lb{background Hamiltonian}
\bold H^{(0)}=NV_o\left(-\frac{\kappa \pi^2_a}{12 a}+\frac{ \pi^2_\phi}{2a^3}+a^3 U\right),
\eq
where $N$ denotes the lapse function and $U$ represents the potential of the scalar field. With the standard Poisson bracket $\{a, \pi_a\}=\{\phi,\pi_\phi\}=1/V_o$, it is straightforward to derive the corresponding Hamilton's equations for the background dynamics which give rise to the classical Friedmann and Raychaudhuri equations in cosmology.

Regarding the scalar perturbations over the homogeneous background, one of the most straightforward approaches to obtaining the Mukhanov-Sasaki equation is to employ the spatially flat gauge in which the perturbations of the metric components are gauged to vanish while the linear perturbation of the scalar field turns out to be the Mukhanov-Sasaki variable. After performing a time-dependent canonical transformation to remove the cross term in the second-order Hamiltonian for the perturbed scalar field and its conjugate momentum,  one can obtain the second-order Hamiltonian for the gauge invariant Mukhanov-Sasaki variable $Q_{\vec k}$, which takes the form \cite{Li:2022evi}
\bq
\lb{Hamiltonian_1}
\bold H^{(2)}=N  V_o \sum_{\substack{\vec k^+}} \left(\frac{|P_{Q_{\vec k}}|^2}{a^3}+a\left(k^2+\Omega^2\right)|Q_{\vec k}|^2\right) .
\eq
Here in order to avoid double counting of the physical degrees of freedom, we sum over only the $\vec k^+$ modes with its first non-vanishing component of the wavevector being strictly positive and
\bq
\lb{potential_1}
\Omega^2=\frac{3\kappa \pi^2_\phi}{a^4}-18\frac{ \pi^4_\phi}{\pi^2_a a^6}-12a\frac{ \pi_\phi U_{, \phi}}{\pi_a}+a^2U_{, \phi \phi},
\eq
where $U_{, \phi}$ and $U_{, \phi \phi}$ stand for the first and second order derivatives of the potential with respect to the scalar field. From the Hamiltonian, one can then derive the Mukhanov-Sasaki equation in terms of $Q_{\vec k}$, yielding
\bq
\lb{ms_equation}
\ddot Q_{\vec k}+3H\dot Q_{\vec k}+\frac{k^2+\Omega^2}{a^2}Q_{\vec k}=0.
\eq
In order to write it as an harmonic oscillator with a time-dependent mass, we need to remove the first order derivative term in the above equation.  This can be achieved by using the rescaled variable $\nu_{\vec k}=aQ_{\vec k}$ whose equation of motion can be shown as
\bq
\lb{ms1}
\nu^{\prime\prime}_{\vec k}+\left(k^2+\Omega^2-\frac{a^{\prime\prime}}{a}\right)\nu_{\vec k}=0,
\eq
where a prime denotes differentiation with respect to the conformal time $d\eta=dt/a$. This defines a time-dependent mass function
\bq
\lb{mass1}
m^2_\mathrm{SF}=\Omega^2-\frac{a^{\prime\prime}}{a},
\eq
where the index `SF' implies the mass function is obtained within the spatially flat gauge. Since equations (\ref{ms_equation}) and (\ref{ms1}) contain the same information of the linear perturbations,  we will also refer to the latter as the Mukhanov-Sasaki equation in the rest part of the paper.

An equivalent form of the Mukhanov-Sasaki equation can be obtained by working in $\nu_{\vec k}$ from the very beginning when a time-dependent canonical transformation is performed. Instead of obtaining the Hamiltonian of $Q_{\vec k}$, one can make a new transformation to arrive at the Hamiltonian of $\nu_{\vec k}$ and then work out its corresponding equation of motion.  These procedures can lead directly to the form (more details can be found in \cite{Li:2022evi})
\bq
\lb{ms2}
\nu^{\prime\prime}_{\vec k}+\left(k^2+\tilde m^2_{\mathrm{SF}}\right)\nu_{\vec k}=0.
\eq
with the mass function given explicitly  by
\bq
\lb{mass2}
\tilde m^2_{\mathrm{SF}}=-\frac{27 \pi^4_\phi}{2\pi^2_aa^6}+\frac{5\kappa \pi^2_\phi}{2a^4}+\frac{9 \pi^2_\phi U}{\pi^2_a}-12aU_{,\phi}\frac{ \pi_\phi}{\pi_a}-\frac{\kappa^2\pi^2_a}{72a^2}+a^2U_{, \phi \phi}-\frac{\kappa}{2}a^2U,
\eq
which is a function of the canonical variables in the phase space. It turns out that two mass functions $ m^2_{\mathrm{SF}}$ and $\tilde m^2_{\mathrm{SF}}$ only differ by a term proportional to the background Hamiltonian constraint and thus they are equivalent on the physical solutions of the classical background dynamics \cite{Li:2022evi}.

In addition to the above two forms of the mass function, its third equivalent form in the classical theory can be obtained from the Einstein-Hilbert action in GR by using the comoving gauge as discussed in detail in \cite{Baumann:2009ds}. In the comoving gauge, the perturbation of the scalar field is set to vanish and the spatial metric up to the linear order in the scalar perturbation can be written as
\bq
g_{ij}=a^2(1-2\mathcal R)\delta_{ij},
\eq
where $\mathcal R$ stands for the comoving curvature perturbation whose magnitude remains constant for the modes outside the comoving  Hubble horizon during inflation. Using the relation between the comoving curvature perturbation and the Mukhanov-Sasaki variable $ \nu_{\vec k}$, namely $\nu_{\vec k}=z_s\mathcal R_{\vec k}$ with $z_s=a\dot{ \phi}/H$,  one can truncate the action to the second order in $\nu_{\vec k}$ and then find the Mukhanov-Sasaki equation in term of $\nu_{\vec k}$, which takes the form \cite{Baumann:2009ds}
\bq
\lb{ms3}
\nu^{\prime \prime}_{\vec k}+\left(k^2-\frac{z^{\prime \prime}_s}{z_s}\right)\nu_{\vec k}=0.
\eq
Thus, we end up with the third form of the mass function, namely,
\bq
\lb{mass3}
m^2_\mathrm{CG}=-\frac{z^{\prime \prime}_s}{z_s},
\eq
here the index `CG' implies that the above mass function is derived in the comoving gauge.
It is straightforward to check that the third form of the classical mass function is also equivalent to the first two forms on the classical trajectories. Although the above three forms of the mass functions are equivalent in the classical dynamics, they can give rise to different effective mass functions that exhibit distinct properties in the Planck regime when they are polymerized in the effective description of LQC. In particular, the first two forms yield respectively the effective mass functions in the dressed metric and the hybrid approaches \cite{ElizagaNavascues:2017avq,Li:2022evi} while the third form can lead to new  forms of the effective mass functions as will be discussed in detail in the next section.

\section{The alternative mass functions in loop quantum cosmology}
\lb{sec:new_mass_function}

In this section, we first briefly review the thumb rules used to promote the first two classical mass functions to the effective mass functions of the Mukhanov-Sasaki equation  in the dressed metric and the hybrid approaches in LQC. Since this part has already been discussed in detail in our earlier paper \cite{Li:2022evi}, we skip some steps in the calculations and refer the reader to above paper. The second part of the section is the gist of the manuscript: based on the third form of the classical mass function given in (\ref{mass3}), a set of alternative effective mass functions are proposed by applying some justifiable polymerization schemes at the level of effective LQC. Some criteria to generate well-behaved effective mass functions throughout the whole evolution of the background are also discussed.

\subsection{The effective mass functions in the dressed metric and the hybrid approaches}
\lb{sec:eff_mass}

In the dressed metric and the hybrid approaches for the linear perturbations in LQC, the classical Hamiltonian constraint is first truncated to the second order in the perturbations and then a hybrid way of quantization is carried out with respect to the background and the linear perturbations. Specifically the background dynamics governed by the homogeneous Hamiltonian constraint (\ref{background Hamiltonian}) is polymerized in the $\bar \mu$ scheme in LQC \cite{Ashtekar:2006wn} while the linear perturbations are Fock quantized. At the level of the effective dynamics, the effective mass functions in the dressed metric and the hybrid approaches can be obtained from their classical counterparts  by using the following substitutions
\bq
\lb{sub_1}
\frac{1}{\pi_a}\rightarrow -\frac{H}{2v^{2/3}\rho},\quad \frac{1}{\pi^2_a}\rightarrow\frac{\kappa}{12v^{4/3}\rho},
\eq
which once plugged into (\ref{mass1}) and (\ref{mass2}) lead to the exact expressions of the effective mass functions in the dressed and hybrid approaches, namely
\bqn
\lb{dressed_mass}
m^2_\mathrm{dressed}&=&-\frac{4\pi G}{3}a^2\rho\left(1+2\frac{\rho}{\rho_c}\right)+4\pi G a^2 P\left(1-2\frac{\rho}{\rho_c}\right)+\mathfrak{U},\\
\lb{hybrid_mass}
m^2_\mathrm{hybrid}&=&-\frac{4\pi G}{3}a^2 \left(\rho-3P\right)+\mathfrak{U},
\eqn
with
\bq
\lb{effective_potential2}
\mathfrak{U}=a^2\left(U_{, \phi \phi}+48 \pi G U+6H\frac{\dot {\phi}}{\rho}U_{, \phi}-\frac{48 \pi G}{\rho}U^2\right).
\eq

These two effective mass functions coincide and tend to the same classical limit in the low curvature regime and their differences become manifest only in the bounce regime. Moreover, it has been found in the previous work \cite{ElizagaNavascues:2017avq} that their difference essentially arises from the non-commutativity of the polymerization and the evaluation of the Poisson brackets. Finally, it is worth pointing out  that the polymerization of $1/\pi_a$ mentioned in (\ref{sub_1}) originates from the hybrid approach. It has been proposed in \cite{Li:2019qzr} to  extend this ansatz to the dressed metric approach since such a polymerization respects the superselection rules for the background quantum dynamics while the original ansatz used in the dressed metric approach does not. However, with regard to the phenomenological investigations, different polymerization ansatz of $1/\pi_a$ would not result in detectable effects if the bounce is dictated by the kinetic energy of inflaton field as is generally the case, as a result of which $\mathfrak{U}$ is  negligible.
As we will show, this negligible role of the  effective potential will be changed when we consider the polymerization of the third form of the classical mass function given in (\ref{mass3}). There the properties of the effective mass function near the bounce also rely on the polymerization ansatz of $1/\pi_a$ even when the bounce is dominated by the kinetic energy of the scalar field.

\subsection{Alternative effective mass functions in LQC}

In the above subsection, we discussed the way polymerization of the mass functions given in (\ref{mass1}) and (\ref{mass2}) can result in the effective mass functions of the Mukhanov-Sasaki equation in the dressed metric and the hybrid approach. Now let us investigate the analogs of the classical mass function (\ref{mass3}) in the effective LQC. Since the classical mass function (\ref{mass3}) is a functional of $z_s(=a\dot { \phi}/H)$, its  polymerization can be implemented by directly
polymerizing  $z_s$. Therefore, as $z_s$ diverges at the quantum bounce where the Hubble rate vanishes, we must polymerize the inverse Hubble rate to make the resulting effective mass function finite at the bounce point.  Here let us note that the effective dynamics of $\bar \mu$ scheme in LQC \cite{Ashtekar:2006uz} can be obtained from the polymerization of the momentum variable $b$  which in the classical regime is proportional to the Hubble rate.\footnote{In the background LQC, the variable $b = c/{\sqrt{|p|}}$ where $c$ denotes the symmetry reduced connection and $p$ denotes the symmetry reduced triad variable which is square of the scale factor (modulo volume of the fiducial cell).  Classical Hamilton's equations yield $c = \gamma \dot a$ where $\gamma$ is the Barbero-Immirzi parameter. Thus, in the classical theory, $b = \gamma H$. In the effective dynamics encoding polymerization terms of connection, this relation gets complicated but surprisingly the modified Friedmann equation takes a simple form: $H^2 = \frac{8 \pi G}{3} \rho\left(1 - \frac{\rho}{\rho_c}\right)$ (for details see for eg. \cite{Ashtekar:2011ni}). }  Therefore, the polymerization of the square of the Hubble rate in the classical mass functions of the Mukhanov-Sasaki equation should be carried out in the same way as for the polymerization of the square of the $b$ squared in the background homogeneous Hamiltonian constraint. While the background effective dynamics guides us directly towards the polymerization of Hubble rate, there are ambiguities when one deals with the inverse Hubble rate and different choices are possible.
Inspired by the ansatz employed to yield the effective mass functions in the dressed metric and the hybrid approaches, in particular the one related with polymerizing $1/\pi_a$ as given in (\ref{sub_1}), we adopt a general ansatz at the phenomenological level with an attempt to exhaust all the possible forms of the alternative  effective mass functions resulting from the polymerization of the classical mass function (\ref{mass3}). The general ansatz takes a very simple form
\bq
\lb{ansatz3}
z_s\rightarrow \frac{a\dot {\phi} f(\rho)}{H},
\eq
with $f(\rho)$ being a function of only energy density that must satisfy the following asymptotic conditions \footnote{Note that in the separate universe approach when applied to LQC, the classical Mukhanov-Sasaki equation still holds for the long-wavelength modes. Hence in principle one can assume $z_s$ can be treated unpolymerized for long wavelength modes and equivalently $f=1$. Although this leads to a divergent $z_s$ at the bounce point, the resulting comoving curvature perturbation is still finite \cite{Wilson-Ewing:2015sfx}. Above assumption  is not valid if one does not make a separate universe approximation. In this manuscript, we discuss the general case beyond this approximation when $z_s$ is polymerized and well behaved in the evolution.}
\bq
\lb{boundary_condition}
\evalat[\Bigg]{\frac{f(\rho)}{\sqrt{\textcolor{blue}{\rho_c-\rho}}}}{\rho= \rho_c}=\mathrm{const},  \quad  \quad \evalat[\Big]{f(\rho)}{\rho\ll \rho_\mathrm{c}}\rightarrow 1,
\eq
here $\rho_c=3/(8\pi G \gamma^2\lambda^2)\approx 0.41$ denotes the maximum energy density in LQC at which the quantum bounce takes place (see footnote 1). Note that this ansatz certainly includes the one used for polymerizing $1/\pi_a$ in (\ref{sub_1}) as a particular case as long as we choose $f=1-\rho/\rho_c$. The first condition in (\ref{boundary_condition}), ensures the finiteness of the polymerized $z_s$ right at the bounce point while the second condition is to recover the classical limit in the low curvature regime.
With the ansatz (\ref{ansatz3}), it is straightforward to show that the classical mass function (\ref{mass3}) can be polymerized into
\bq
\lb{effective_mass_3}
 m^2_\mathrm{z}=\tilde{\mathfrak{U}}-\frac{a^{\prime \prime}}{a},
\eq
where
\bq
\lb{effective_potential}
\tilde{\mathfrak{U}}=a^2\left(U_{, \phi \phi}+(48 \pi G+\delta_a)U+(6H\frac{\dot {{\phi}}}{\rho}+\delta_b)U_{, \phi}+(\delta_c-\frac{48 \pi G}{\rho})U^2+\delta_d \rho^2\right),
\eq
with the correction terms given explicitly by
\bqn
\lb{correction_term_a}
 \delta_a&=&\frac{192 \pi G \rho^2 f_{,\rho \rho}}{f}\left(1-\frac{\rho}{\rho_c}\right)+\frac{48 \pi G \rho f_{,\rho}}{f}\left(1+\frac{\rho}{\rho_c}\right)-\frac{96 \pi G \rho^2}{\rho_c(\rho-\rho_c)},\\
 \lb{correction_term_b}
\delta_b&=&-12H\dot {\phi}\frac{f_{,\rho}}{f}-\frac{6 H \dot {\phi}}{\rho_c-\rho},\\
\lb{correction_term_c}
\delta_c&=&-\frac{96 \pi G \rho f_{,\rho \rho}}{f}\left(1-\frac{\rho}{\rho_c}\right)+\frac{48 \pi G  f_{,\rho}}{f}\left(1-2\frac{\rho}{\rho_c}\right)-\frac{48 \pi G (2\rho-\rho_c)}{\rho_c(\rho_c-\rho)},\\
\lb{correction_term_d}
\delta_d&=&-\frac{96 \pi G \rho f_{,\rho \rho}}{f}\left(1-\frac{\rho}{\rho_c}\right)-\frac{48 \pi G  f_{,\rho}}{f}\left(2-\frac{\rho}{\rho_c}\right)+\frac{48 \pi G }{\rho-\rho_c}.
\eqn
It is important to note that the $a^{\prime \prime}/a$ term in the effective mass (\ref{effective_mass_3}) should be evaluated on the trajectories of the effective dynamics in LQC. Therefore, when all the correction terms  vanish, the above effective potential (\ref{effective_potential}) reduces to the one given in (\ref{effective_potential2}). Correspondingly, the effective mass function (\ref{effective_mass_3}) reduces to the one in the dressed metric approach. However, it can be shown that one can not find a viable $f$ satisfying (\ref{boundary_condition}) to make all the correction terms vanish at the same time. To be specific, from the condition $\delta_b=0$, one can solve for  $f$ which turns out to be
\bq
f=\sqrt{1-\frac{\rho}{\rho_c}}.
\eq
Substituting the above expression into $\delta_a$, one can immediately find that
\bq
\lb{delta_a}
\delta_a=-\frac{24\pi G \rho}{\rho_c}.
\eq
As a result, demanding $\delta_a=\delta_b=0$ at the same time is impossible,  implying that the effective mass function (\ref{effective_mass_3}) can not reduce to the one used in the dressed metric approach. Similarly, in order to obtain the effective mass function in the hybrid approach from (\ref{effective_mass_3}), we need
\bq
\delta_a=-\frac{16 \pi G \rho}{\rho_c},\quad \quad \delta_b=0,\quad \quad \delta_c=0,\quad \quad\delta_d=\frac{32\pi G}{3\rho_c},
\eq
which is also impossible since for a vanishing $\delta_b$, the value of $\delta_a$ has to be given by (\ref{delta_a}).  As a result, any feasible effective mass functions from (\ref{effective_mass_3}) would be different from the one used in either the dressed metric or the hybrid approach.

Therefore, by choosing different forms of $f$, one can end up with other alternative effective mass functions in the Mukhanov-Sasaki equation of LQC. Moreover, it is important to note that the correction  terms listed in (\ref{correction_term_a})-(\ref{correction_term_d}) are not all dependent on the potential of the scalar field. In particular, the last term $\delta_d\rho^2$ can become dominant in the Planck regime. As a result, unlike in the dressed metric approach or the hybrid approach, the effective potential (\ref{effective_potential}) can still play an important role in the Planck regime even for the kinetic-energy-dominated bounce as long as $\delta_d$ does not vanish. On the other hand, with a vanishing $\delta_d$, the predictions on the observations by using  our new effective mass function will be expected to be of no substantial difference from those obtained in the dressed metric approach when the bounce is dominated by the kinetic energy density of the scalar field. Therefore in the following we are more interested to search for new effective mass functions that have non-vanishing $\delta_d$ with  well-behaved correction terms throughout the non-singular evolution of the background spacetime.

As expected the asymptotic condition (\ref{boundary_condition}) and the finiteness of the correction terms throughout the whole evolution impose very strict restrictions on the possible forms of the function $f$. Firstly, it should be noted that in the expressions of the correction terms (\ref{correction_term_a})-(\ref{correction_term_d}) there are individual terms which will become divergent right at the quantum bounce. Therefore, a reasonable choice of $f$ should make all these correction terms finite at the bounce and meanwhile satisfy the asymptotic condition given in (\ref{boundary_condition}). Further analysis becomes more manageable if we start with the $\delta_b$ term given in (\ref{correction_term_b}). In order to get rid of  the divergence  at $\rho=\rho_c$ which originates from the second term on the right-hand side of (\ref{correction_term_b}), the only feasible choice of $f$ turns out to be
\bq
f=\sqrt{\rho_c-\rho}~g(\rho),
\eq
where $g$ can in principle be any well-behaved function of the energy density which satisfies the following asymptotic conditions
\bq
\lb{boundary_condition_2}
\evalat[\Big]{g(\rho)}{\rho=\rho_c}=\mathrm{const}
\quad  \quad \mathrm{and} \quad  \quad  \evalat[\Big]{g(\rho)}{\rho\ll \rho_\mathrm{c}}\rightarrow \frac{1}{\sqrt{\rho_c}}.
\eq
The simplest choice of $g$ which satisfies the above condition is
\bq
\lb{ansatz1}
g_1=\frac{1}{\sqrt{\rho_c}}.
\eq
This ansatz corresponds to the original polymerization ansatz of $1/\pi_a$ which was used in the original dressed metric approach initially proposed in \cite{Agullo:2012fc}. It can lead to the following correction terms
\bq
\lb{alternative_mass1}
\delta_{a_1}=-24 \pi G \frac{\rho}{\rho_c},\quad \quad \delta_{b_1}=0,\quad\quad \delta_{c_1}= \frac{24 \pi G}{\rho_c},\quad \quad \delta_{d_1}=0,
\eq
where the subscripts $a_1$, $b_1$, $c_1$ and $d_1$ refer to the choice $g_1$, and these correction terms are finite in the whole range $\rho\in[0,\rho_c]$. Although it gives rise to the first alternative effective mass function other than those used in the dressed metric and the hybrid approach, due to a vanishing $\delta_d$, we expect the corresponding effective potential is negligible in the bouncing regime for the kinetic-dominated bounce. As a result, this new effective mass function is not very interesting to us.

Our second ansatz stems from the hybrid approach. In particular, the polymerization of $1/\pi_a$ prescribed in (\ref{sub_1}) corresponds to the choice
\bq
\lb{ansatz2}
g_2=\frac{1}{\sqrt{\rho_c}}\left(1-\frac{\rho}{\rho_c}\right)^{1/2},
\eq
which results in the new correction terms
\bq
\lb{alternative_mass2}
\delta_{a_2}=-48 \pi G \frac{\rho}{\rho_c},\quad \quad \delta_{b_2}=\frac{6H \dot \phi}{\rho-\rho_c},\quad\quad \delta_{c_2}=0 ,\quad \quad \delta_{d_2}=\frac{48\pi G}{\rho_c},
\eq
where the subscripts $a_2$, $b_2$, $c_2$ and $d_2$ refer to the choice $g_2$.
One can then immediately find from the above results that the second ansatz fails to yield a viable effective mass function due to the divergence of $\delta_{b_2}$ at the bounce point. Besides, we find that as long as the form of the function $g$ is chosen to be $g=\frac{1}{\sqrt{\rho_c}}\left(1-\frac{\rho}{\rho_c}\right)^{\alpha},$ then the only viable value of $\alpha$ that can result in finite correction terms in the range $\rho\in[0,\rho_c]$ is $\alpha=0$. Moreover, it turns out that this result can also be extended to a more general setting where the function $g$ takes the form
\bq
g(\rho)=\frac{1}{\sqrt{\rho_c}}\left(1-\frac{\rho}{\rho_c}\right)^{\alpha}\tilde g(\rho),
\eq
with $\tilde g(\rho)$ being a differentiable function of the energy density and meanwhile satisfying the condition $\tilde g(\rho_c)\neq 0$. With this ansatz, it can be shown that in order to make $\delta_b$ finite at the bounce point, the only possible choice is still $\alpha=0$. Thus, we conclude that in order to ensure a non-divergent $\delta_b$ at the bounce point, the necessary condition is $g(\rho)=\tilde g(\rho)/\sqrt{\rho_c}$ with $g(\rho)$ non-vanishing. Correspondingly, in terms of $g(\rho)$, the resulting correction terms turn out to be
\bqn
\delta_a&=&-\frac{24 \pi G \rho}{\rho_c  g }\left( g-2\left(\rho_c-3\rho\right)  g'+8 \rho\left(\rho-\rho_c\right){ g}^{\prime\prime}\right), \\
\delta_b&=&-12 H \dot{\phi}\frac{ g'}{ g}, \\
\delta_c&=&\frac{24 \pi G}{\rho_c  g }\left( g+2\rho_c  g'+4 \rho\left(\rho-\rho_c\right){g}^{\prime\prime}\right), \\
\delta_d&=& -\frac{48 \pi G}{\rho_c  g }\left(\left(2\rho_c-3\rho\right)  g'+2 \rho\left(\rho_c-\rho\right){ g}^{\prime\prime}\right),
\eqn
where a prime denotes differentiation with respect to the energy density.
Since $ g(\rho_c)\neq 0$, the above correction terms are all finite at the bounce point. The exact form of   $g(\rho)$ is directly related with the construction of the quantum theory, in particular, the way to polymerize the inverse momentum $1/b$ (or equivalently $1/\pi_a$). For example, if the inverse momentum is polymerized as
\bq
\lb{b1}
\frac{1}{b}\rightarrow \frac{\lambda\cos^{\epsilon}\left(\lambda b\right)}{\sin\left(\lambda b\right)},
\eq
with $\epsilon=0$ corresponding to the first case mentioned in (\ref{ansatz1}) and $\epsilon=1$ to the second case in (\ref{ansatz2}),
then the inverse Hubble rate is correspondingly polymerized as 
\bq
\lb{inverse_hubble_rate}
\frac{1}{H}=\frac{\gamma}{b}\rightarrow \frac{\lambda\gamma\cos^{\epsilon}\left(\lambda b\right)}{\sin\left(\lambda b\right)}=\frac{\cos^{\epsilon+1}\left(\lambda b\right)}{H}=\frac{(1-\rho/\rho_c)^{(\epsilon+1)/2}}{H}.
\eq
In the above derivation, on the left-hand side of the arrow we have used the relation between the Hubble rate and the momentum $b$  in the classical theory which is $b=\gamma H$  while on the right-hand side of the arrow we have used the relations valid in the effective dynamics, namely, $
H=\sin\left(\lambda b\right)\cos\left(\lambda b\right)/(\lambda \gamma)$ and $\rho=\rho_c\sin^2(\lambda b)$. Hence, given the ansatz  (\ref{b1}), the function $f$ is given by $f=(1-\rho/\rho_c)^{(\epsilon+1)/2}$ and the corresponding $g(\rho)$ turns out to be 
\bq
\lb{gtilde_1}
g(\rho)=\frac{1}{\sqrt{\rho_c}}\left(1-\frac{\rho}{\rho_c}\right)^{\epsilon/2}.
\eq
Combining with the requirement $ g(\rho_c)\neq 0$, we can conclude that with regard to the polymerization ansatz (\ref{b1}), only $\epsilon=0$ can result in well-behaved correction terms and thus is the only feasible choice. The resulting correction terms are exactly those given in (\ref{alternative_mass1}).

In principle, other ways of polymerization can lead to different forms of $g(\rho)$. A correct polymerization should originate from the cosmological sector of LQG. Since it is still an open question to extract the cosmological sector from full LQG, here we present a phenomenological example to illustrate the possibility that the well-behaved correction terms with a non-vanishing $\delta_d$ can arise when the inverse momentum is properly polymerized. For example, we can consider the following  polymerization ansatz
\bq
\lb{b2}
\frac{1}{b}\rightarrow \frac{\lambda}{\sin\left(\lambda b\right)}\Big[1+\xi \sin^2\left(\lambda b \right)\Big],
\eq
where the parameter $\xi$ satisfies the condition $\xi>-1$. Following the procedure presented in (\ref{inverse_hubble_rate}), it is straightforward to show that the above ansatz leads to the choice
\bq
\lb{gtilde_2}
g(\xi)=\frac{1}{\sqrt{\rho_c}}\left(1+\xi \frac{\rho}{\rho_c}\right),
\eq
which in turn results in the following correction terms in the effective mass function
\bqn
\delta_a&=&-\frac{24\pi G\rho}{\rho_c+\xi\rho}\left(1-2\xi+7\xi\frac{\rho}{\rho_c}\right),\quad \delta_b=-12 H \dot{ \phi}\frac{\xi}{\rho_c+\xi\rho}, \nb\\
\delta_c&=&\frac{24\pi G}{\rho_c+\xi\rho}\left(1+2\xi+\xi\frac{\rho}{\rho_c}\right),\quad \delta_d=-\frac{48 \pi G \xi}{\rho_c\left(\rho_c+\xi\rho\right)}\left(2\rho_c-3\rho\right).
\eqn
Therefore, as long as $\xi>-1$, all the correction terms are finite and well-behaved in the interval $\rho\in[0,\rho_c]$. Besides, for a large range of the parameter space of $\xi$, $\delta_d$ is non-zero. Moreover, one can estimate the magnitude of the effective mass function near the bounce. When the bounce is dominated by the kinetic energy of the scalar field, only the $\delta_d$ term in the potential (\ref{effective_potential}) plays the dominant role. As a result, near the bounce the effective mass function can be approximated by
\bq
m^2_\mathrm{z}\approx-\frac{4\pi G}{3}a^2\rho\left(1+2\frac{\rho}{\rho_c}\right)+4\pi G a^2 P\left(1-2\frac{\rho}{\rho_c}\right)+a^2\delta_d \rho^2,
\eq
where the contributions from the potential are ignored.
In particular, right at the bounce, the effective mass function becomes
\bq
\evalat[\Big]{m^2_\mathrm{z}}{\rho=\rho_c}\approx 8\pi G a^2\rho_c\left(\frac{5\xi-1}{1+\xi}\right).
\eq
Therefore, the sign of the effective mass function at the bounce depends on the magnitude of the parameter $\xi$. From this example, we can see the property of the alternative effective mass function in the Planck regime is closely related with the polymerization of the inverse momentum $1/b$ (or equivalently $1/\pi_a$). This is a novel property of the alternative effective mass function given in (\ref{effective_mass_3})-(\ref{correction_term_d}) as compared with those in the dressed metric and the hybrid approach.

\section{Conclusions}
\lb{sec:summary}

In this manuscript, we have derived a general form of the alternative effective mass functions, given explicitly by Eqs. (\ref{effective_mass_3})-(\ref{effective_potential}), in the Mukhanov-Sasaki equation of LQC by applying a general polymerization ansatz of the inverse Hubble rate in the classical expression of $z_s$. It originates from an equivalent form of the classical mass function which differs from those used in the dressed metric and the hybrid approach. Our investigation starts with a brief review of the Hamiltonian formulation of the linear perturbation theory in cosmology with the purpose of presenting two forms of the classical mass functions which after polymerization are promoted respectively to the effective mass functions used in the dressed metric and the hybrid  approach.  The third form of the classical mass function given in (\ref{mass3}) can be obtained from the action by using the comoving gauge. Although these three forms of the classical mass function turn out to be equivalent on the trajectories of the classical Friedmann dynamics, they actually lead to the distinct effective mass functions when the same \emph{hybrid} way of quantization is implemented, namely, the background dynamics is loop quantized while the perturbations are Fock quantized. The modified Mukhanov-Sasaki equation for all these approaches is obtained by polymerizing the background quantities in its classical counterpart.

The quantum gravity effects in the Mukhanov-Sasaki equation of LQC are taken into account by polymerizing the classical mass functions. At the level of the effective dynamics, polymerization of the classical mass function is implemented by polymerizing the  terms related with $1/\pi_a$ and $1/\pi^2_a$ in the dressed metric and the hybrid approach. When it comes to the third form of the classical mass function, based on $z_s$ (\ref{mass3}), only the polymerization of $1/\pi_a$ is required. Thus, the properties of the alternative effective mass functions resulting from (\ref{mass3}) are closely related with the polymerization ansatz of $1/\pi_a$. From our analysis we have shown that the classical mass function (\ref{mass3}) can not lead to the effective mass function in either the dressed metric approach or the hybrid approach, implying that (\ref{mass3}) can only result in the alternative forms of the effective mass functions. In particular, the polymerization of (\ref{mass3}) can give rise to additional four correction terms in the effective potential (\ref{effective_potential}) which are denoted respectively by $\delta_a$, $\delta_b$, $\delta_c$ and $\delta_d$. Among them,  the correction terms $\delta_a$ and $\delta_b$ can not vanish at the same time. Moreover, for the bounce which is dominated by the kinetic energy of the scalar field, only the $\delta_d$ term will become important in the Planck regime which has the potential to yield different predictions other than those in the dressed metric or the hybrid approach. Combined with the criteria that the effective mass function should be well-behaved throughout the whole evolution and meanwhile tend to its classical counterpart in the low curvature regime, we have in particular tested three polymerization ansatz of $1/\pi_a$ (or equivalently inverse momentum $1/b$ or inverse Hubble rate) which are employed in the original dressed metric approach, in the hybrid approach and the one given as a phenomenological model in (\ref{b2}). It turns out that the first ansatz can yield well-behaved effective mass function in the whole range $\rho\in[0,\rho_c]$ but with a vanishing $\delta_d$ which makes this ansatz less interesting since it is expected to lead to the similar predictions as those in the dressed metric approach. On the other hand, the second ansatz from the hybrid approach can only result in an ill-behaved $\delta_b$ term which becomes divergent right at the quantum bounce and thus fails to work at all. Finally, our last ansatz successfully leads to well-behaved effective mass function with a non-vanishing $\delta_d$. Its properties in the Planck regime near the quantum bounce is completely determined by a phenomenological parameter $\xi$. The main problem with such an ansatz is lack of understanding of its quantum origin. In principle, polymerization of $1/\pi_a$ should come from the cosmological sector of full LQG which is currently unavailable. Finally, it is worth noting that the difference between our new alternative effective mass functions and those used in the dressed metric and the hybrid approaches is due to the non-commutativity between the evaluation of the Poisson brackets and the polymerization procedures. As discussed in our previous paper \cite{Li:2022evi}, the same reason results in the different effective mass functions used in the dressed metric and the hybrid approach.

In summary, our investigation shows that the precise form of the effective mass function for perturbations in  LQC is far from completely settled. Due to quantization ambiguities, different forms of the classical mass functions which are equivalent in the classical dynamics can be polymerized into physically distinct effective mass functions once quantum gravity effects are taken into account. The effective mass functions used in the dressed metric and the hybrid approach are just two examples of these effective mass functions. More alternatives to these two mass functions are possible to be discovered when equivalent forms of the classical mass functions are polymerized in certain ways. Moreover, the phenomenological impacts of these effective mass functions on the CMB observations are far from being thoroughly explored. When matching the CMB observations, one should numerically investigate the full parameter space of the model before drawing any conclusions on the feasibility of the different perturbation approaches and this is one of the topics worthy of future investigations.

\section*{Acknowledgments}

We thank Martin Bojowald and Mikhail Kagan for a comment and Edward Wilson-Ewing for useful discussions. B.-F. Li is supported by the National Natural Science Foundation of China (NNSFC) with the grant No. 12005186. P.S. is supported by NSF grant PHY-2110207.

\end{document}